\documentclass[prl,aps,showpacs,floatfix,twocolumn]{revtex4-1}
\usepackage{color}
\usepackage{colordvi}
\usepackage{graphicx,amssymb,epsfig,amsmath}
\usepackage{epstopdf}
\usepackage{amsmath}
\usepackage{bm}
\usepackage{amsfonts}

\begin{document}
    
    \title{Order parameter and detection for crystallized dipolar bosons in lattices}
    \author{Budhaditya Chatterjee}
    \email{bchat@iitk.ac.in}
    \affiliation{Department of Physics, Indian Institute of Technology-Kanpur, Kanpur 208016, India}
    \author{Axel U. J. Lode}
    \email{axel.lode@univie.ac.at}
    \affiliation{Wolfgang Pauli Institute c/o Faculty of Mathematics, University of Vienna, Oskar-Morgenstern Platz 1, 1090 Vienna, Austria}
    \affiliation{Vienna Center for Quantum Science and Technology, Atominstitut, TU Wien, Stadionallee 2, 1020 Vienna, Austria}
    \affiliation{Department of Physics, University of Basel, Klingelbergstrasse 82, CH-4056 Basel, Switzerland}

    \begin{abstract}
        We explore the ground-state properties of bosons with dipole-dipole interactions in a one-dimensional optical lattice. For strong interactions, the interesting phenomenon of crystallization takes place. Herein, we provide a detailed characterization and a way to measure the resulting crystal phase. Using the eigenvalues of the reduced one-body density matrix we define an order parameter that yields a phase diagram in agreement with an analysis of the density and two-body density. We demonstrate that the phase diagram can be detected experimentally using the variance of single-shot measurements.
    \end{abstract}
    
    \maketitle
    
    Dipolar ultracold atoms have attracted much interest recently~\cite{baranov08,lahaye09}. This interest is corroborated by experimental realizations of dipolar Bose-Einstein condensates (BECs) of chromium~\cite{griesmaier05,beaufils08}, dysprosium~\cite{lu11}, and erbium~\cite{aikawa12} atoms as well as potassium-rubidium molecules~\cite{ni08}. The realization of ultracold sodium-potassium molecules with tunable dipolar interaction strength enables the exploration of a regime where dipolar interactions are dominant~\cite{zwierlein15}. 
    
    The long-range and anisotropic dipole-dipole interaction leads to rich and exotic quantum effects distinct from the effects of BECs with contact interactions. Dipolar bosons in a trap show elongation of the condensate along the direction of the oriented dipoles~\cite{yi01, santos00, goral00}. Counterintuitively, the stability of dipolar condensates increases in special geometries like very oblate traps~\cite{yi01, santos00, eberlein05, goral02a,koch08}. The long-range and anisotropic interactions make dipolar ultracold atoms a great resource to explore quantum phases and aspects of many-body physics~\cite{lahaye09,goral02,menotti07,zoller10}.
    
    In contrast to condensed matter setups, the parameters of cold-atom systems can be controlled almost at will. Changing the dimensionality may yield additional features such as $p$-wave superfluidity in two-dimensional Fermi gases~\cite{bruun08, cooper09}, Luttinger-liquid-like behavior in one-dimensional bosons~\cite{arkhipov05,citro07,depalo08,pedri08} and an anisotropy effect for bosons on a ring~\cite{zollner11,zollner11pra,maik11}. 
    
    Even few-particle systems can be deterministically produced in experiments~\cite{selim1} and enable the investigation of the fundamental building blocks of many-body systems, for instance, in lattices~\cite{selim2,selim3} from a bottom-up perspective. Moreover, these few-atom systems can be handled numerically accurately with any precision for any inter-particle interaction strength~\cite{lode12,axel1} allowing the investigations of properties emergent for strong interactions, like, for instance, fermionization in lattices~\cite{brouzos10,alon05}. 
    
    Ultracold atoms in lattices can serve as a quantum simulator for condensed matter systems~\cite{landig15,jotzu14,goldman16,gross17}. Dipolar atoms in optical lattices, due to their long-range anisotropic interactions, have an enriched phase diagram as compared to systems with contact interactions: a density wave phase~\cite{goral02, dalla06} and Haldane insulating phases~\cite{dalla06, deng11} were predicted in one- and two-dimensional systems. 
    
    In this work, we focus on a remarkable property that renders strongly interacting dipolar systems significantly different from atoms with contact interactions: a crystallization process is seen for one-dimensional homogeneous dipoles~\cite{arkhipov05}, dipoles in a linear and a zigzag chain~\cite{astrakharchik08a,astrakharchik08b}, in a harmonic trap~\cite{deuretzbacher10}, a ring geometry~\cite{zollner11,zollner11pra}, and in a triple well~\cite{chatterjee12}.  
    
    Theoretically, dipolar atoms in triple wells have, for instance, been investigated using mean-field methods~\cite{peters12}, the Hubbard model~\cite{lahaye10,anna13,fischer13,gallemi13,gallemi16}, and using the multi-configurational time-dependent Hartree (MCTDH) method~\cite{chatterjee12}. The physics of strong dipolar interactions lie beyond the area of validity of mean-field methods and standard Hubbard models~\cite{zollner11,zollner11pra,chatterjee12,cao17,tan16,Streltsov13,lu12}; the usage of a general many-body approach is necessary. We follow the strategy of Refs.~\cite{selim1,selim2,selim3} and find the many-body system's physics from analysis and understanding of its few-body building blocks. We solve the few-body Schr\"odinger equation for the ground states of bosons with dipole-dipole interactions in a potential with a few wells using the MCTDH for bosons (MCTDHB) method~\cite{alon08} implemented in the MCTDH-X software~\cite{ultracold,axel1,axel2}. Chiefly, we investigate a triple well because it is the elemental building block that exhibits all essential long-range dipolar effects of bosons in optical lattices.
    
    We theoretically investigate the ground-state phase diagram  as a function of the strength of dipole-dipole interactions and as a function of the depth of the lattice potential for commensurate fillings. We provide a detailed characterization and measurement protocol for the result of the crystallization process -- the crystal phase of dipolar bosons. We analyze the one-body and two-body density and identify the characteristic density signatures for all the emergent phases, the superfluid (SF), the Mott-Insulator (MI), and the crystal state (CS). We demonstrate how the natural populations, i.e., the eigenvalues of the reduced one-body density matrix, can be used to identify distinct phases. We define and analyze an order parameter that is a function of these natural populations. This order parameter yields a phase diagram that can be straightforwardly measured using the variance of images obtained from single-shot measurements. 

    We consider $N$, dipolar bosons in a quasi-one-dimensional optical lattice with all their dipoles polarized along the same direction. 
    \begin{equation}
        {
            H=\sum_{i=1}^{N}-\frac{\hbar^{2}}{2M}\partial_{x_{i}}^{2}+\sum_{i=1}^{N}V_{tw}(x_{i})+\sum_{i<j} V_{int}(x_i-x_j)
        }.
        \label{Eq.Ham}
    \end{equation}
    
    The optical lattice, $V_{ol}= V\sin^{2}(\kappa x)$ has the depth $V$ and the wave vector $\kappa$. We consider $S$ lattice sites with hard wall boundary conditions. A strong transverse confinement of characteristic length $a_\perp$  prevents any excitation into the transverse direction and ensures the quasi-one-dimensionality. 
    
    In the absence of s-wave scattering, the quasi-one-dimensional dipole-dipole interaction is modeled as $V_{int}(x_i-x_j) = g_c\delta(x_i-x_j) + \frac{g_d}{\vert x_i - x_j\vert^3 + \alpha}$ ($g_c,g_d$ are the  coupling constants). For large separations $\vert x_i - x_j\vert\gg a_\perp$, we get the far-field long-range dipole-dipole interaction $\sim 1/r^3$. For small separations $|x_i - x_j|\lessapprox a_\perp$, the transverse confinement regularizes the divergence at $x_i=x_j$ by introducing an effective interaction cutoff $\alpha \approx {a_\perp}^3$ and an additional $\delta$-like-interaction $g_c\delta(x_i-x_j)$~\cite{sinha07,deuretzbacher10,cai10}. The coupling constants are related as $g_c\approx g_d \frac{4\sqrt{\gamma}}{3\sqrt{2\pi}}$, where $\gamma$ is the trap aspect ratio $\dfrac{a_\parallel}{a_\perp}$~\cite{cai10}. 
    We henceforth refer to $g_d$  as the strength of the interaction and is given as $g_d = d^2_m /4\pi\epsilon_0$ for electric dipoles and as $g_d = d^2_m \mu_0 /4\pi$ for magnetic dipoles, where $d_m$ is dipole moment, $\epsilon_0$ the vacuum permittivity, and $\mu_0$ the vacuum permeability.
    
     To arrive at convenient dimensionless units, we rescale the Hamiltonian in Eq.~\eqref{Eq.Ham} by the lattice recoil energy $E_{R}=\hbar^{2}\kappa^{2}/2M$ and investigate a triple well setup ($S=3$) with hard wall boundaries at $x=\pm S\pi/2\kappa$, an aspect ratio $\gamma = 25.6$, and an interaction cutoff $\alpha=0.05$. Results for other parameters are shown in the Supplementary Information~\cite{SI}. Since commensurability is necessary to realize the aforementioned phases, we will discuss exclusively commensurate fillings.
    
    \begin{figure}
        \centering
        \includegraphics[width=1.0\linewidth]{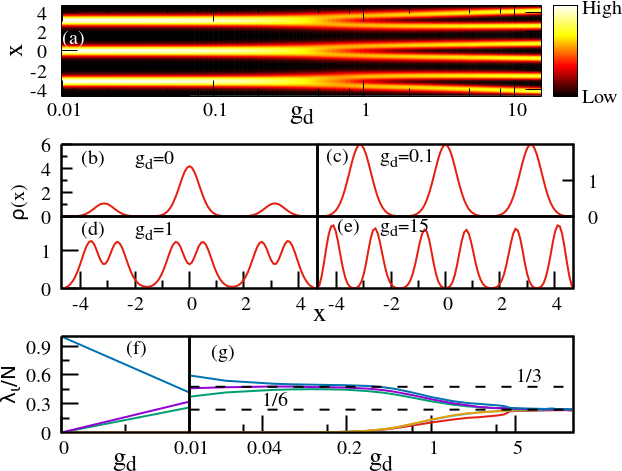}
        \caption {(a) The one-body density $\rho(x)$ for $N=6$ bosons as a function of the strength of the dipolar interactions $g_d$. The depth of the lattice is fixed at $V=8$. The density shows a transition from a threefold to a sixfold spatial splitting as the dipolar interaction strength increases. (b)--(e) One-body densities representative for different phases. For the superfluid phase (b) the kinetic and trap energies dominate: the density is maximal in the central well. For the Mott-Insulator phase (c) the bosons are equally distributed between all wells. In (d) the onset of fermionization and formation of the Tonks gas is mimicked by the formation of a characteristic dip at $x=0$ and in (e) the emergence of well-separated density peaks is one of the hallmarks of the crystal phase. (f)--(g) Normalized population of natural orbitals as a function of the interaction strength $g_d$. At $g_d\approx0$, only the first orbital has a significant population. As the interaction increases, $n=3$ orbitals begin to populate reaching an equal population in the Mott-Insulator state. For large $g_d$ as many orbitals as there are particles ($N=6$) are populated, and the system reaches the maximally fragmented crystal state.}     
        \label{fig:den1d}
    \end{figure}
    
    For bosons with contact interactions in a lattice, the competition between the kinetic and interaction energy determines the quantum phases~\cite{greiner02,bruder}. For dipolar atoms, however, this competition is changed~\cite{chatterjee15,streltsov13}: the relative strengths of the short-ranged portion and the long-ranged portion of the interaction affect the ground-state properties in addition to the kinetic energy. This is particularly important at large dipolar interaction strengths $g_d$~\cite{comment2}. 

    Let us clarify our adoption of the terminology ``phases'' and ``order parameter''. Rigorously, the concept of quantum phases and, likewise, their order parameter is legitimate solely in the thermodynamic limit. The finite size ensemble we consider cannot exhibit true macroscopic phases in this strict sense. Nonetheless, for small ensembles, the ground-state possesses properties that are analogous to the macroscopic phases. These analogous properties are the ``finite-size precursors'' to the macroscopic quantum phase \cite{Luhmann08}. In our work, finite-size effects are still present but do not dictate the physics; for the sake of simplicity, we thus use the terms ``phase'' and ``order-parameter'' when discussing the finite-size precursors of the quantum phase.
        
    We first analyze the one-body density of the ground state as a function of the repulsive dipolar interaction strength $g_d$ for a fixed depth $V$ of the lattice [Fig.~\ref{fig:den1d}(a)]. 
    For small interactions $g_d\approx0$, the kinetic energy dominates the competition of energies and the bosons are in a coherent superposition of all number-states (i.e states counting the occupation of each lattice site). This fully delocalized state represents the superfluid phase. Dominating kinetic energy and hard-wall boundary conditions lead to a maximal population in the central well [Fig.~\ref{fig:den1d}(b)]. 
    
    As the interaction strength increases, the short-range portion of the interaction begins to dominate, and the bosons localize in each well. The density exhibits a single maximum in each well. This Mott-Insulator phase is characterized by the localization of atoms in the lattice with vanishing overlap of the densities in distinct wells, Fig.~\ref{fig:den1d}(c).
    
    When the interaction strength increases further, the short-range interactions become strong enough for the bosons to attempt to fermionize; a characteristic dip in the center of the one-body density in each well emerges [Fig.~\ref{fig:den1d}(d)]. Fermionization refers to systems resembling a Tonks gas for which the density is identical to the density of non-interacting fermions~\cite{girardeau60}. The original Bose-Fermi map is valid only for contact interactions. For dipolar interactions, a similar mapping can be constructed by exploiting the divergence of the interaction potential when the positions of two atoms are equal~\cite{astrakharchik08b,comment1}.

    In the strongly interacting limit, the long-ranged $\sim1/r^3$ tail of the dipolar interaction becomes dominant and determines the physics. As a consequence of the long-range interactions well-separated density maxima for each particle in the system emerge [Fig.~\ref{fig:den1d}(e)]; this localized structure of the density is a feature of the crystal phase and marks the departure of the physics of the system from the area of validity of the Hubbard model.
    
    \begin{figure}
        \centering
        \includegraphics[width=1\linewidth]{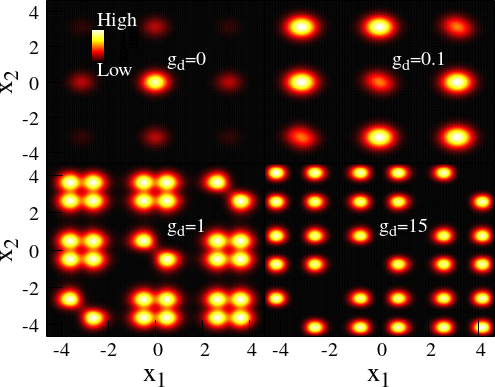}
        \caption{Exploring the two-body density in the emergent phases. In the absence of dipole-dipole interactions ($g_d=0$), the atoms cluster at $(x_1=x_2=0)$. The delocalization of the atoms in the superfluid causes the non-central peaks to locate at $x_1=0,x_2=\pm\pi$ and $x_1=\pm\pi,x_2=0$ while the central peak is at $x_1=x_2=0$. At $g_d=0.1$, the localization of atoms in the Mott-insulating phase is evident from the depletion of the diagonal of the two-body density. At $g_d=1.0$, as the dipoles fermionize as a ``correlation hole'' develops: the two-body density tends to zero for $x_1=x_2$. At
            $g_d=15$, the dipoles crystallize and the two-body density forms a square lattice with missing diagonal. }
        \label{fig:den2d}
        
    \end{figure}
    
    Next, we illustrate the mechanism of the localization process for the different phases using the two-body density $\hat{\rho}_2(x_1,x_2)= \langle \Psi \vert \hat{\Psi}^{\dagger}(x_1)\hat{\Psi}^{\dagger}(x_2) \hat{\Psi}(x_1) \hat{\Psi}(x_2) \vert \Psi \rangle$ (Fig.~\ref{fig:den2d}). 
    As interactions increase, the transition from the superfluid to the Mott state (Fig.~\ref{fig:den2d}, upper row) is seen from the change in $\rho_2$ from a maximum at center $x_1=x_2=0$ to off-diagonal $x_1 \neq x_2$ maxima and a depletion of its diagonal: the atoms start to minimize the probability to be at the same position in space whilst remaining in the lowest band of the lattice [Fig.~\ref{fig:den2d}($g_d=0.1$)]. With a further increase of the interaction strength, $\rho_2$ develops a correlation hole, i.e., $\rho_2(x,x)\rightarrow0$, implying that the probability of finding two bosons in the same place is reduced [Fig.~\ref{fig:den2d}($g_d=1.0$)]. This resembles the behavior of hard-core bosons with infinitely strong contact interactions [cf. Fig.~\ref{fig:den1d}(d)].  
    
    Eventually, when the dipolar interaction strength is increased beyond the Tonks regime [Fig.~\ref{fig:den2d}($g_d=15$)], the long-range tail of the interaction determines the physics of the system: the two-body density shows a complete spatial isolation of every particle in a square-lattice-like pattern with a missing diagonal; a hallmark of the transition from the fermionized gas to the crystal state. The crystal state is a pure long-range interaction effect and cannot be reached with contact interactions alone [cf. Fig.~\ref{fig:den2d}($g_d=15$), Fig.~7 in Ref.~\cite{brouzos10}]. 
    
    Next, we show that the eigenvalues of the reduced one-body density matrix or natural populations can be used to define an order parameter that identifies all the phases of dipolar atoms in optical lattices.
    
    The reduced one-body density matrix is defined as
    \begin{equation}
        \hat{\rho}_1(x,x')=\langle \Psi \vert \hat{\Psi}^{\dagger}(x) \hat{\Psi}(x') \vert \Psi \rangle=\sum_i \lambda_i \varphi^*_i(x) \varphi_i(x^{\prime}).\label{RDM}
    \end{equation}
    The second equality illustrates that $\rho_1$ can be diagonalized to yield an expansion in terms of its eigenfunctions $\varphi_i(x)$, and its eigenvalues $\lambda_i$, also termed natural orbitals and occupations, respectively.
    
    If only one natural population is macroscopic, the system is condensed~\cite{penrose56}, and if several natural populations are macroscopic, the system is said to be fragmented~\cite{spekkens,mueller}. 
    
    For small interactions, the bosons are condensed and form a superfluid; only the lowest natural orbital is populated and only the first natural population is macroscopic, $\lambda_1 \approx N$, Fig.~\ref{fig:den1d}(f). 
    
    With the transition to the Mott-insulating phase, fragmentation emerges: the reduced one-body density matrix attains as many equally large eigenvalues as there are lattice sites, while all other eigenvalues are zero, Fig.~\ref{fig:den1d}(g). In a system of $N$ particles in $S$ sites, the significant eigenvalues of the reduced density matrix are hence equal to $N/S$: $\lambda_i\approx N/S$ for $i\leq S$ and $\lambda_i\approx0$ for $i>S$. 
    
    As interactions increases beyond the Mott-insulating phase, more natural orbitals become populated until fragmentation is maximal and the crystal state forms: $N$ orbitals attain unit population, irrespective of the number of sites $S$, Fig.~\ref{fig:den1d}(g).
    
    In order to quantify our above observations using the one-body and two-body densities and the natural occupations with a phase diagram, we define the order parameter 
    \begin{equation}
        \Delta= \sum_k \left(\frac{\lambda_k}{N}\right)^2, \label{OP}
    \end{equation}
    
    here $\lambda_k$ is the $k^{th}$ natural occupation, cf. Eq.~\eqref{RDM}. For the superfluid phase only one eigenvalue $\lambda_1$ is non-negligible and, hence, $\Delta=1$. For the Mott-Insulator as many eigenvalues as there are sites in the lattice are contributing equally, while the rest is negligible and $\Delta=\frac{1}{S}$. The crystal state is characterized by $\Delta=\frac{1}{N}$; each of the $N$ bosons occupies a separate orbital, while the other occupations $\lambda_k$ with $k>N$ are negligible.
    
    We stress that the TG gas is an intermediate stage to the MI$\rightarrow$CS transition. While there exists a universal behavior of the natural occupations for the SF, MI and the CS phases, there is none for the TG gas. Hence, unlike in the case of the SF, MI and CS states, there is no value of $\Delta$ associated with the TG gas.

    \begin{figure}
        \centering
        \includegraphics[width=1\linewidth]{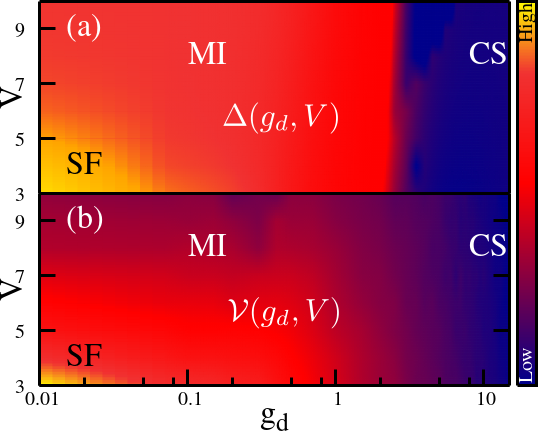}
        \caption{(a) Phase diagram showing the order parameter $\Delta(g_d,V)$ as a function of the lattice depth $V$ and the dipolar interaction strength $g_d$. The values of $\Delta$ for the Superfluid (SF), the Mott-Insulator (MI), and for the Crystal State (CS) are, respectively $\Delta\approx 1$, $\Delta\approx 1/3$, and $\Delta\approx 1/6$. The SF phase can be observed for a small barrier heights $V$ and low interaction strength $g_d$. As $V$ increases, the SF$\rightarrow$MI transition occurs. As $g_d$ increases, the MI$\rightarrow$CS transition takes place. The CS is a maximally fragmented state which forms as a result of dominating long-range interactions. Unlike the SF$\rightarrow$MI transition, the MI$\rightarrow$CS transitions is almost independent of the barrier height because the crystal transition is a genuine many-body effect.
            (b) Simulation of the phase diagram using the variance of single-shots in real-space $\mathcal{V}(g_d,V)$. For every point, the variance of $10000$ single-shot images were computed. The value of the variance $\mathcal{V}$ is maximal for the SF phase, decreases in the MI phase and attains its minimum value for the CS. The phase diagram using $\mathcal{V}(g_d,V)$ shows all the three phases and closely resembles the one obtained using the order parameter $\Delta(g_d,V)$.}
        \label{fig:Vdphase}
    \end{figure}
    
    We now discuss the phase diagram of dipolar bosons obtained by plotting the order parameter $\Delta$ as a function of the barrier height $V$ and the interaction strength $g_d$, Fig.~\ref{fig:Vdphase}. The superfluid phase is restricted to small values of $V$ and $g_d$. The Mott-Insulator emerges for increasing dipolar interactions. 
    The occurrence of maximal fragmentation marks the emergence of the crystal phase where the short-ranged contribution of the dipolar interactions saturates while the long-ranged contribution~\cite{comment2} of the dipolar interaction potential forces the bosons to become fully separated and fragmented. 
    
    A fundamental difference between the transition SF$\rightarrow$MI and the transition MI$\rightarrow$CS is the mechanism of fragmentation. In the SF$\rightarrow$MI transition, fragmentation is \textit{extrinsic}, because it is governed by the one-body lattice potential of the Hamiltonian. Here, the eigenvalues of the reduced density matrix are dependent on the number of lattice sites. In the transition MI$\rightarrow$CS, fragmentation is \textit{intrinsic}, because it is governed exclusively by the dipolar two-body interaction in the Hamiltonian and is not dependent on the one-body potential. Here, the eigenvalues of the reduced density matrix do not depend on the number of lattice sites. The crystal state formation is thus a genuine many-body effect. This explains why the crystallization is seen also in other one-dimensional systems, irrespective of boundary conditions or the one-body potential~\cite{arkhipov05,deuretzbacher10,zollner11,zollner11pra,astrakharchik08a}.
    
    While we explicitly present the result for a few-body system, we stress that our order-parameter is absolutely general and valid for any \textit{large but finite} system, like the ones produced in cold-atom experiments.
    In the thermodynamic limit  $N \rightarrow \infty$ and $S \rightarrow \infty$, we find $\Delta \rightarrow 0$ for, both, the MI and CS  phase. However, $\Delta$ approaches zero with different rates $\tau$, determined by $N/S$ ratio, for the MI and the CS cases.
    While the order parameter $\Delta$ does coincide for the MI and CS phases in the thermodynamic limit, the finite-size scaling of the rate $\tau$ at which $\Delta$ approaches zero with the system size ($\frac{N}{S} \neq 1$) still reveals if the system is in the CS or the MI state. 
    
    While the order parameter $\Delta$ reflects each quantum phase, the detection of $\Delta$ or the eigenvalues $\lambda_k$ of the reduced one-body density matrix remains an experimental challenge. Further, the direct measurement of the densities of the system with a resolution sufficient to detect intra-site features also represents a formidable problem; this problem can be solved since the lattice constant is experimentally tunable and intra-well structures are thus resolvable~\cite{oberthaler06}.
    Single-shot measurements contain information about the correlations of the atoms in Bose-Einstein condensates~\cite{sakmann16}. In particular, the variance of single-shot measurements in momentum space has been found to yield valuable information on the natural occupations~\cite{lode17}. A single-shot measurement represents a sample of the positions of all $N$ particles drawn from the $N$-particle probability distribution associated with the many-body wave-function $\vert \Psi \rangle$ (see Supplementary Information~\cite{SI}).
    
    In the following, we discuss how to detect the phase diagram [Fig.~\ref{fig:Vdphase}(a)] using single-shot measurements.
    Here, we adopt a similar approach to the one taken in Ref.~\cite{lode17} and compute the variance of simulations of single-shot measurements in real space. We use spatial instead of momentum measurements because the spatial single-shot distributions let us access the degree of localization of the bosons. In the present system of dipolar bosons in a lattice, the degree of localization increases with increasing dipolar interaction strength [Fig.~\ref{fig:Vdphase}(b)]: in the superfluid, the atoms are delocalized in the entire lattice. The single-shot variance $\mathcal{V}$ is largest for this state since the atoms can be sampled from any position in the lattice. In the Mott-Insulator, the atoms localize in individual lattice sites while they are delocalized within each lattice site. As a consequence, the variance in the single-shot measurements decreases as compared to the superfluid. In the crystal phase, the atoms localize in individual natural orbitals, i.e., they form a structure irrespective of the underlying lattice potential. This further localization through maximizing the fragmentation further minimizes the variance in single-shot measurements. In summary, we find that the single-shot variance $\mathcal{V}$ matches the phase diagram obtained in Fig.~\ref{fig:Vdphase}(a) closely; importantly, the crystal state is clearly discernible using the single-shot variance $\mathcal{V}$. Since we investigate a small system, the phase transitions are not expected to be sharp. Further, the phase transitions of the system are smoothened in single-shot measurements as compared to the order parameter $\Delta$, because in the transitions between the phases in a finite system, the orbitals which are delocalized between sites become occupied. Since our simulations of single-shot measurements correspond to standard absorption imaging, we thus have shown a straightforward way for experimental verification of our findings.

    \acknowledgments{BC gratefully acknowledges the financial support from Department of Science and Technology, Government of India under DST Inspire Faculty fellowship. AUJL acknowledges financial support by the Swiss SNF and the NCCR Quantum Science and Technology, the Austrian Science Foundation (FWF) under grant No. F65 (SFB ``Complexity in PDEs''), and the Wiener Wissenschafts- und TechnologieFonds (WWTF) project No MA16-066 (``SEQUEX''). Computation time on the Hazel Hen cluster of the HLRS in Stuttgart and the HPC2013 cluster of the IIT Kanpur are gratefully acknowledged. BC thanks AK Agarwal, IITK for help with an initial computational resource.}


\begin{thebibliography}{100}
\bibitem{comment2} For a lattice potential, the ``short-range part of the dipolar interaction'' refers to the part of the interaction potential that is confined within a single site; it can be quantified using an integral of the interaction potential over one site, $V_{\rm S} \equiv \int_0^{\frac{\pi}{2}} {\frac{g_d}{x^3 + \alpha}} dx$.
        The ``long-range part of the interaction'' refers to the part of the interaction potential in the domain excluding the local site; it can be quantified using an integral of the interaction potential on all but the local site, $V_{\rm L} \equiv \int_{\frac{\pi}{2}}^{\infty} {\frac{g_d}{x^3 + \alpha}} dx$.
\bibitem{comment1} The exact mapping is valid for $\alpha = 0$ exploiting the divergence at $x_i = x_j$. Since in our case $\alpha \neq 0$, the mapping is not valid in a strict sense. For sufficiently large $g_d$, the fermionization effect can, however, be seen clearly in our results.
\end{thebibliography}
\end{document}


\title{Supplementary Information\\Order parameter and detection for crystallized dipolar bosons in lattices} 
\author{Budhaditya Chatterjee}
\email{bchat@iitk.ac.in}
 \affiliation{Department of Physics, Indian Institute of Technology-Kanpur, Kanpur 208016, India}
\author{Axel U. J. Lode}
\affiliation{Wolfgang Pauli Institute c/o Faculty of Mathematics, University of Vienna, Oskar-Morgenstern Platz 1, 1090 Vienna, Austria}
\affiliation{Department of Physics, University of Basel, Klingelbergstrasse 82, CH-4056 Basel, Switzerland}
\affiliation{Vienna Center for Quantum Science and Technology, Atominstitut, TU Wien, Stadionallee 2, 1020 Vienna, Austria}
\email{axel.lode@univie.ac.at}
\maketitle

The main text discusses the emergence of different phases and phase diagram of ultracold dipolar bosons and how to measure it using the variance in simulations of single shot measurements. The results in the main text are for $S=3$ sites and $N=6$ bosons.
In this Supplementary Information, Section~\ref{params} shows the generality of the results in the main text for larger lattices and different filling factors.
Sec.~\ref{var_SS} introduces the background of single-shot simulations and how to compute the variance analyzed in Fig.~3b) of the main text.
Sec.~\ref{Mom} shows and analyzes the momentum distribution.

 \section{Results for different system parameters}\label{params}

In this section we show the Fig.~1 results of the main text for larger lattice size $S=5$ as well as larger filling fraction $\nu=3$ to demonstrate that the results obtained in the main text can be generalized. In a comparison of the results in the main text to results without the short-ranged contribution, $g_c \delta(x-x')$, we confirmed that no qualitative differences are triggered by the inclusion of the additional short-ranged contribution. We therefore neglect the additional short-ranged contribution $g_c \delta(x-x')$ in this section.

\subsection{Different lattice sizes}

\begin{figure}
	\centering
	\includegraphics[width=1.0\linewidth]{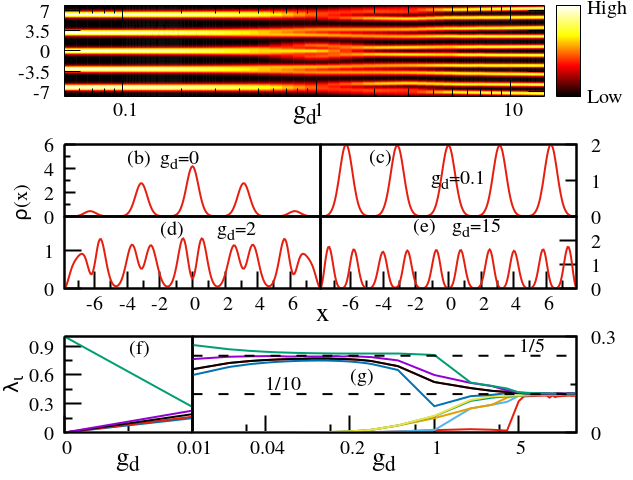}\vspace*{-4mm}
	\caption{Results of Fig.~1 for larger lattice size ($S=5$).
		Top panel: One-body density $\rho(x)$ for $S=5, N=10$ plotted as a function of space $x$ and dipolar coupling $g_d$ for $V=8$. For small $g_d$ the density exhibits a fivefold splitting corresponding to the $S=5$ lattice sites as we are in MI state. As the dipolar interaction strength $g_d$ increases, a transition from a fivefold spatial splitting to a tenfold spatial splitting corresponding to $N=10$ bosons takes place as we transition to the crystal state. (b)--(e) One-body densities representative for different phases. (b) $g_d=0$, the superfluid phase with a maximal population at the center. (c) $g_d=0.1$, the Mott-insulating phase. Five peaks corresponds to a localization in each well with each site having two particles. (d) $g_d=2.0$, fermionization and mimicking of the Tonks gas can be seen from the characteristic dips in the density within each well. (e) $g_d=15.0$, the crystal state formation can be seen from the emergence of well-separated density peaks as the long-ranged interaction starts to dominate.
		(f)--(g) Normalized population of natural orbitals as a function of the interaction strength $g_d$. The superfluid ($g_d\approx0$) is condensed and only one orbital contributes significantly. For the Mott insulator as many orbitals as there are sites ($S=5$) contribute; all having an equal population of $\approx 0.2$ .  For large $g_d$ as many orbitals as there are particles ($N=10$) contribute with equal population of $\approx 0.1$ and the system reaches the maximally fragmented crystal state.
	}
	\label{fig:den1dS5N10}
\end{figure}

To demonstrate the generality of our results for three wells and six particles in the main text also for larger lattices, we show results for five lattice sites ($S=5$) and ten particles ($N=10$) in Fig.~\ref{fig:den1dS5N10}. 
Similar to the main-text results, the one-body density evolves from an initial superfluid phase to a Mott-Insulator phase, characterized by five separate density peaks (corresponding to $S=5$ lattice sites). Stronger interactions result in intra-well splitting of the bosons as a result of the repulsive dipolar interaction. Finally, at the crystal phase, ten separated density peaks are seen corresponding to $N=10$ particles.

It is thus seen that the conjectures of the main text are valid also for larger lattices with commensurate filling.

\subsection{Different filling factors}

\begin{figure}
	\centering
	\includegraphics[width=1.0\linewidth]{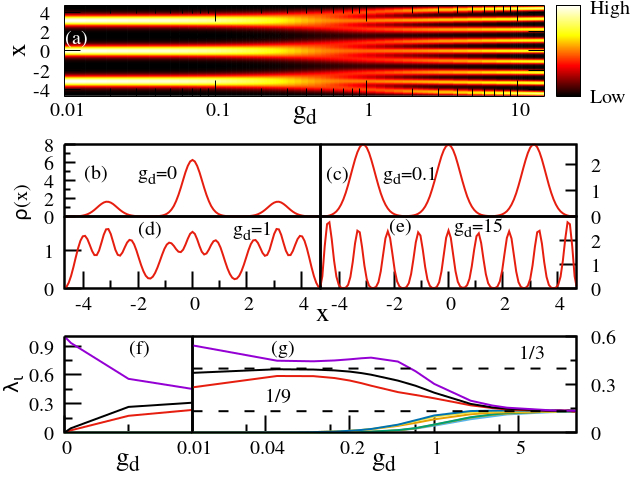}\vspace*{-4mm}
	\caption{Results of Fig.~1 for higher filling factor ($S=3,N=9$). 
		Top panel: One-body density $\rho(x)$ for $S=3, N=9$ plotted as a function of space $x$ and dipolar coupling $g_d$ for $V=8$. For small $g_d$ in the superfluid and Mott-insulating phases the density is split threefold, because the lattice has three sites ($S=3$).
		Since the filling factor is $\nu=3$ with a total of $N=9$ bosons, the density splits into three parts at each well forming an overall ninefold split structure in the crystal phase, as interactions become stronger. (b)--(e) One-body densities representative for different phases. (b) $g_d=0$, the superfluid with a maximal population in the central well. (c) $g_d=0.1$, the Mott-insulator shows localization in each site forming three density peaks. (d) $g_d=1.0$, the mimicking of the Tonks gas is seen from the emergence of characteristic dips that result in a three-hump structure within every site.
		(e) $g_d=15.0$, the crystal state formation can be seen from the emergence of well-separated density peaks for each particles as the long-ranged interaction starts to dominate.
		(f)--(g) Normalized population of natural orbitals as a function of the interaction strength $g_d$. In the superfluid phase ($g_d\approx0$), the system is condensed and only one orbital contributes significantly. For the Mott-insulator as many orbitals as there are sites ($S=3$) contribute, all having approximately equal populations of $\approx 0.33$. For large $g_d$, in the crystal phase, as many orbitals as there are particles ($N=9$) are populated equally having population $\approx 0.11$.
	}
	\label{fig:den1dS3N9}
\end{figure}

To demonstrate the generality of our results in the main text for two particles per lattice site also for different commensurate fillings, we show results for a triple-well lattice ($S=3$) with nine particles ($N=9$), i.e., a filling of three particles per site in Fig.~\ref{fig:den1dS3N9}.

Here, the transition from the SF to MI phase is exhibited with three density peaks corresponding to $S=3$ lattice sites. With increasing interaction, since each lattice site is triply occupied, we see a three-fold splitting in each well. At the crystal phase, the splitting is complete and we obtain nine density peaks corresponding to $N=9$.

It is thus seen that the findings documented in the main text are true also for different commensurate filling factors.

\clearpage

\section{Measuring the phase diagram of ultracold dipolar atoms with single shots}\label{var_SS}
Experimental absorption images measure the positions of all particles simultaneously. To simulate such an absorption image, we use the wave-functions $\Psi(x_1,...,x_N)$ obtained from our MCTDHB simulations as a starting point. A single absorption image corresponds to drawing all particle positions $(s_1,...,s_N)$ simultaneously from the probability $\vert \Psi(x_1,...,x_N)\vert^2$. We use the algorithm documented in Refs.~\cite{SingleShots,lode17} to draw $N_{\text{shots}}$ samples from a given state $\Psi(x_1,...,x_N)$. For every simulated single shot measurement, i.e., every single sample which is drawn, a convolution with a three pixels wide Gaussian is performed to emulate the point spread function of a realistic imaging system. The result of these convolutions are $N_{\text{shots}}$ functions $\lbrace \mathcal{B}_j (x) \rbrace_{j=1}^{N_\text{shots}}$. From these functions, the single-shot variance $\mathcal{V}$ is obtained as follows: 
\begin{equation}
  \mathcal{V}= \int dx \frac{1}{N_\text{shots}} \sum_{j=1}^{N_\text{shots}} \left[ \mathcal{B}_j(x) - \bar{\mathcal{B}}(x) \right]^2;\qquad \bar{\mathcal{B}}(x) = \frac{1}{N_\text{shots}} \sum_{j=1}^{N_\text{shots}} \mathcal{B}_j(x)\:.
\end{equation}

\section{Momentum distribution}\label{Mom}

In this section analyze the one-particle momentum density:

\begin{equation}
\tilde{\rho}(k)=2\pi \langle k|\rho_1|k \rangle =\int dx \int dx' e^{-ik(x-x')} \rho^{1}(x,x').
\end{equation}

\begin{figure}
	\centering
	\includegraphics[width=0.8\linewidth]{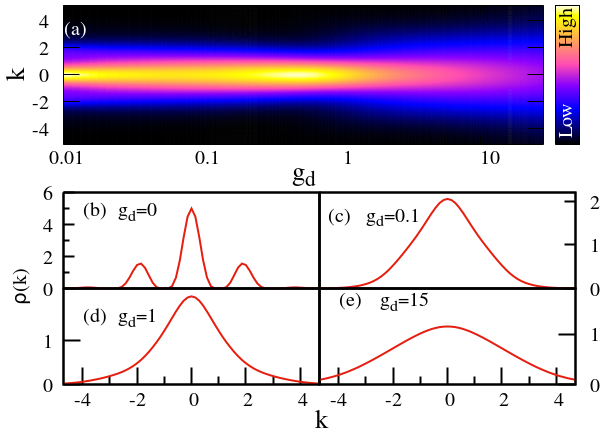}
	\caption{Tracing the evolution of the momentum density as a function of dipole-dipole interaction strength. (a) Momentum density profile $\tilde{\rho}(k)$ as a function of interaction strength $g_d$. The initially centrally peaked momentum density shows spreading as the strength of the dipole-dipole interactions increase. (b)--(e) Shows $\tilde{\rho}(k)$ at different interaction strengths characteristic to the emergent phases [cf. Fig.~1(b)--(e) in the main text]. At $g_d=0$ there is a central sharp peak, with smaller peaks near the reciprocal lattice vectors. With increasing $g_d$, the small side peaks are suppressed while the central peak becomes flattened. The crystal state at $g_d=15$ features a central peak that is strongly spread to a plateau-like structure.}
	\label{fig:den1dk}
\end{figure}

Fig.~\ref{fig:den1dk} shows the momentum density $\tilde{\rho}$ as a function of $k$ and $g_d$ for $V=8$. For very weak interactions $g_d \approx 0$, the momentum density has a central sharp peak at $k=0$, with additional smaller peaks on the side. This is due to the fact that since the bosons are spatially delocalized over the lattice: the momentum distribution (which is the fourier transform of density), consequently, are \textit{localized} and show Bragg peaks near the reciprocal lattice vectors $k=2\pi/a$ ($a$ being the lattice spacing).

As the strength of the interactions increases, the increasing spatial localization of the bosons in the MI phase manifests in the smearing of the central peak and suppression of the Bragg peaks. This demonstrates the reduction of coherence in the MI phase. Further increasing the strength of interactions triggers no qualitative change in $\tilde{\rho}$. The only quantitative change is a spreading of $\tilde{\rho}$, indicating further spatial localization and corresponding momentum delocalization as the system reaches the crystal state. 

We note here, the absence of a clear signature of the phase transition MI$\rightarrow$CS in the momentum distributions $\tilde{\rho}$. Hence, the momentum distribution is not a good candidate to observe the phase transition to the crystal phase.